\documentstyle[12pt]{article}

\def\hybrid{\topmargin 0pt      \oddsidemargin 0pt
        \headheight 0pt \headsep 0pt
        \textwidth 17.2cm
        \textheight 24cm
        \voffset=-1.7cm
        \hoffset=-0.4cm
        \marginparwidth 0.0in
        \parskip 5pt plus 1pt   \jot = 1.5ex}
\catcode`\@=11
\def\marginnote#1{}

\newcount\hour
\newcount\minute
\newtoks\amorpm
\hour=\time\divide\hour by60
\minute=\time{\multiply\hour by60 \global\advance\minute by-\hour}
\edef\standardtime{{\ifnum\hour<12 \global\amorpm={am}%
        \else\global\amorpm={pm}\advance\hour by-12 \fi
        \ifnum\hour=0 \hour=12 \fi
        \number\hour:\ifnum\minute<10 0\fi\number\minute\the\amorpm}}
\edef\militarytime{\number\hour:\ifnum\minute<10 0\fi\number\minute}

\def\draftlabel#1{{\@bsphack\if@filesw {\let\thepage\relax
   \xdef\@gtempa{\write\@auxout{\string
      \newlabel{#1}{{\@currentlabel}{\thepage}}}}}\@gtempa
   \if@nobreak \ifvmode\nobreak\fi\fi\fi\@esphack}
        \gdef\@eqnlabel{#1}}
\def\@eqnlabel{}
\def\@vacuum{}
\def\draftmarginnote#1{\marginpar{\raggedright\scriptsize\tt#1}}

\def\draft{\oddsidemargin -0.1truein
        \def\@oddfoot{\sl preliminary draft \hfil
        \rm\thepage\hfil\sl\today\quad\militarytime}
        \let\@evenfoot\@oddfoot \overfullrule 3pt
        \let\label=\draftlabel
        \let\marginnote=\draftmarginnote
   \def\@eqnnum{{\rm (\theequation)}\rlap{\kern\marginparsep\tt\@eqnlabel}%
\global\let\@eqnlabel\@vacuum}  }

\font\teneuf=eufm10  scaled  1440 
\font\seveneuf=eufm7 scaled  1\@ptsize00 
\font\fiveeuf=eufm5  scaled  1\@ptsize00 

\newfam\euffam
\textfont\euffam=\teneuf  \scriptfont\euffam=\seveneuf
  \scriptscriptfont\euffam=\fiveeuf

\def\hexnumber@#1{\ifnum#1<10 \number#1\else
 \ifnum#1=10 A\else\ifnum#1=11 B\else\ifnum#1=12 C\else
 \ifnum#1=13 D\else\ifnum#1=14 E\else\ifnum#1=15 F\fi\fi\fi\fi\fi\fi\fi}

\def\got{\ifmmode\let\next\got@\else
 \def\next{\errmessage{Use \string\got\space only in math mode}}\fi\next}
\def\got@#1{{\got@@{#1}}}
\def\got@@#1{\fam\euffam#1}

\newfont{\lgot}{eufm10 scaled 1440}%

\newfont{\Bbb}{msbm10 scaled 1\@ptsize00}
\newcommand{\CC}{\mbox{\Bbb C}}
\newcommand{\NN}{\mbox{\Bbb N}}

\newcommand{\RR}{\mbox{\Bbb R}}

\newcommand{\ZZ}{\mbox{\Bbb Z}}
\newfont{\Bbbb}{msbm7 scaled 1\@ptsize00}
\newcommand{\z}{\raise-1pt\hbox{$\mbox{\Bbbb Z}$}}
\newcommand{\r}{\raise-1pt\hbox{$\mbox{\Bbbb R}$}}
\font\teneufm=cmmib10 scaled 1\@ptsize00
\font\seveneufm=cmmib7 scaled 1\@ptsize00
\font\fiveeufm=cmmib5  
\def\bfit#1{{\textfont1=\teneufm\scriptfont1=\seveneufm
\scriptscriptfont1=\fiveeufm
\mathchoice{
\hbox{$\mathsurround=0pt\displaystyle#1$}}
{\mathsurround=0pt\hbox{$\textstyle#1$}}
{\hbox{$\mathsurround=0pt\scriptstyle#1$}}
{\hbox{$\mathsurround=0pt\scriptscriptstyle#1$}}}}

\font\sevenmsa=msam6 
\newfam\msafam
\textfont\msafam=\sevenmsa
\def\hexnumber@#1{\ifnum#1<10 \number#1\else
\ifnum#1=10 A\else\ifnum#1=11 B\else\ifnum#1=12 C\else
\ifnum#1=13 D\else\ifnum#1=14 E\else\ifnum#1=15 F\fi\fi\fi\fi\fi\fi\fi}
\def\msa@{\hexnumber@\msafam}
\def\llcorner{\delimiter"4\msa@78\msa@78 }
\def\lrcorner{\delimiter"5\msa@79\msa@79 }
\mathchardef\blacktriangleright="3\msa@49
\mathchardef\blacktriangleleft="3\msa@4A
\mathchardef\trianglerighteq="3\msa@44
\mathchardef\trianglelefteq="3\msa@45
\font\tenmsb=msbm10 scaled 1\@ptsize00
\newfam\msbfam
\textfont\msbfam=\tenmsb
\def\msb@{\hexnumber@\msbfam}
\mathchardef\varkappa="0\msb@7B

\newdimen\linethick  \linethick=0.4pt
\newdimen\hboxitspace    \hboxitspace=5pt
\newdimen\vboxitspace    \vboxitspace=5pt

\def\fr#1{%
\beq\new
\vcenter{
\hrule height\linethick
           \hbox{\vrule width\linethick
                 \kern\hboxitspace
                 \vbox{\kern\vboxitspace
                       \hbox{$\begin{array}{c}\displaystyle#1
          \end{array}$}%
                       \kern\vboxitspace}%
                 \kern\hboxitspace
                 \vrule width\linethick}%
           \hrule height\linethick}%
\eeq}

\newdimen\Squaresize \Squaresize=14pt
\newdimen\Thickness \Thickness=0.5pt

\def\Square#1{\hbox{\vrule width \Thickness
   \vbox to \Squaresize{\hrule height \Thickness\vss
      \hbox to \Squaresize{\hss#1\hss}
   \vss\hrule height\Thickness}
\unskip\vrule width \Thickness}
\kern-\Thickness}

\def\Vsquare#1{\vbox{\Square{$#1$}}\kern-\Thickness}

\def\numberbysection{\@addtoreset{equation}{section}
        \def\theequation{\thesection.\arabic{equation}}}
\numberbysection

\renewcommand{\theequation}{\thesection.\arabic{equation}}
\newcommand{\l@qq}[2]{\addvspace{2em}
 \hbox to\textwidth{\hspace{1em}\bf #1 \dotfill #2}}

\newcounter{app}

\def\app{\setcounter{equation}{0}
\def\theequation{\Alph{app}.\arabic{equation}}\par
   \addvspace{4ex}
   \@afterindentfalse
  \secdef\@app\@dapp}
\newcommand\@app{\@startsection {app}{1}{0ex}%
                                   {-3.5ex \@plus -1ex \@minus -.2ex}%
                                   {2.3ex \@plus.2ex}%
                                   {\normalfont\Large\bf}}

\def\@dapp#1{%
{\parindent \z@ \raggedright  \bf #1}\par\nobreak}
\def\l@app#1#2{\ifnum \c@tocdepth >\z@
    \addpenalty\@secpenalty
    \addvspace{1.0em \@plus\p@}%
    \setlength\@tempdima{2.5em}%
    \begingroup
      \parindent \z@ \rightskip \@pnumwidth
      \parfillskip -\@pnumwidth
      \leavevmode \bfseries
      \advance\leftskip\@tempdima
      \hskip -\leftskip
      #1\nobreak\hfil \nobreak\hb@xt@\@pnumwidth{\hss #2}\par
    \endgroup\fi}
\newcounter{sapp}[app]

\def\sapp{\def\theequation{\Alph{app}.\arabic{equation}}\par
   \@afterindentfalse
  \secdef\@sapp\@dsapp}
\newcommand\@sapp{\@startsection{sapp}{2}{\z@}%
                                     {-3.25ex\@plus -1ex \@minus -.2ex}%
                                     {1.5ex \@plus .2ex}%
                                     {\normalfont\large\bfseries}}

\def\@dsapp#1{%
{\parindent \z@ \raggedright  \bf #1}\par\nobreak}
\newcommand{\l@sapp}{\@dottedtocline{2}{1.5em}{3em}}

\def\titlepage{\@restonecolfalse\if@twocolumn\@restonecoltrue\onecolumn
     \else \newpage \fi \thispagestyle{empty}\c@page\z@
        \def\thefootnote{\fnsymbol{footnote}} }

\def\endtitlepage{\if@restonecol\twocolumn \else  \fi
        \def\thefootnote{\arabic{footnote}}
        \setcounter{footnote}{0}}  
\relax

\hybrid

\newtoks\@stequation

\def\subequations{\refstepcounter{equation}%
  \edef\@savedequation{\the\c@equation}%
  \@stequation=\expandafter{\theequation}
  \edef\@savedtheequation{\the\@stequation}
  \edef\oldtheequation{\theequation}%
  \setcounter{equation}{0}%
  \def\theequation{\oldtheequation\alph{equation}}}

\def\endsubequations{%
  \setcounter{equation}{\@savedequation}%
  \@stequation=\expandafter{\@savedtheequation}%
  \edef\theequation{\the\@stequation}%
  \global\@ignoretrue}

\makeatletter
\newdimen\normalarrayskip              
\newdimen\minarrayskip                 
\normalarrayskip\baselineskip
\minarrayskip\jot
\newif\ifold             \oldtrue            \def\new{\oldfalse}
\def\arraymode{\ifold\relax\else\displaystyle\fi} 
\def\eqnumphantom{\phantom{(\theequation)}}     
\def\@arrayskip{\ifold\baselineskip\z@\lineskip\z@
     \else
     \baselineskip\minarrayskip\lineskip1\baselineskip\fi}

\def\@arrayclassz{\ifcase \@lastchclass \@acolampacol \or
\@ampacol \or \or \or \@addamp \or
   \@acolampacol \or \@firstampfalse \@acol \fi
\edef\@preamble{\@preamble
  \ifcase \@chnum
     \hfil$\relax\arraymode\@sharp$\hfil
     \or $\relax\arraymode\@sharp$\hfil
     \or \hfil$\relax\arraymode\@sharp$\fi}}

\def\@array[#1]#2{\setbox\@arstrutbox=\hbox{\vrule
     height\arraystretch \ht\strutbox
     depth\arraystretch \dp\strutbox
     width\z@}\@mkpream{#2}\edef\@preamble{\halign \noexpand\@halignto
\bgroup \tabskip\z@ \@arstrut \@preamble \tabskip\z@ \cr}%
\let\@startpbox\@@startpbox \let\@endpbox\@@endpbox
  \if #1t\vtop \else \if#1b\vbox \else \vcenter \fi\fi
  \bgroup \let\par\relax
  \let\@sharp##\let\protect\relax
  \@arrayskip\@preamble}
\def\eqnarray{\stepcounter{equation}%
              \let\@currentlabel=\theequation
              \global\@eqnswtrue
              \global\@eqcnt\z@
              \tabskip\@centering                      
              \let\\=\@eqncr
              $$%
            \halign to \displaywidth  \bgroup
             \eqnumphantom \@eqnsel
      \hskip\@centering                               
    $\displaystyle  \tabskip\z@ {##}$%
    &\global\@eqcnt\@ne \hskip 2\arraycolsep
         $ \displaystyle  \arraymode{##}$\hfil
    &\global\@eqcnt\tw@ \hskip 2\arraycolsep
         $\displaystyle\tabskip\z@{##}$\hfil
         \tabskip\@centering
    &{##}\tabskip\z@\cr}
\makeatother
\newtheorem{th}{Theorem}[section]     

\newtheorem{con}{Conjecture}[section]
\newtheorem{lem}{Lemma}[section]

\newtheorem{rem}{Remark}[section]
\def\bea{\begin{eqnarray}}
\def\eea{\end{eqnarray}}

\def\beq{\begin{equation}}
\def\eeq{\end{equation}}
\def\be{\beq\new\begin{array}{c}}  
\def\ee{\end{array}\eeq}           
\def\bse{\begin{subequations}}                
\def\ese{\end{subequations}}                 %

\def\square{\hfill{\vrule height6pt width6pt            
depth1pt} \break \vspace{.01cm}}                        

\def\bgamma{{\bfit\gamma}}

\def\brho{{\bfit\rho}}
\def\al{\alpha}
\def\la{\lambda}

\def\e{\epsilon}
\def\<{\langle}
\def\>{\rangle}
\def\ov{\overline}
\def\wt{\widetilde}
\def\wh{\widehat}
\def\N{\scriptscriptstyle N}
\def\ch{{\cal H}}
\begin{document}

\begin{titlepage}

\begin{center}
\hfill ITEP-TH-54/99\\

\hfill hep-th/9910265\\
\phantom.
\bigskip\bigskip\bigskip\bigskip\bigskip\bigskip
{\Large\bf Integral representation for the eigenfunctions of quantum

\bigskip\noindent
periodic Toda chain}\\
\bigskip \bigskip
{\large S. Kharchev\footnote{E-mail:  kharchev@vitep5.itep.ru}, D.
Lebedev\footnote{E-mail:  dlebedev@vitep5.itep.ru}}\\ \medskip {\it
Institute
of Theoretical \& Experimental Physics\\ 117259 Moscow, Russia}\\
\bigskip \bigskip 
\end{center}

\bigskip
\bigskip
\bigskip
\begin{abstract}
\noindent
Integral representation for the eigenfunctions of quantum periodic
Toda chain is constructed for $N$-particle case. The multiple integral
is calculated using the Cauchy residue formula. This
gives the representation which reproduces the particular results
obtained by Gutzwiller for $N=2,3$ and $4$-particle chain. Our
method to solve the problem combines the ideas of Gutzwiller and
$R$-matrix approach of Sklyanin with the classical results in the theory of
the Whittaker functions. In particular, we calculate Sklyanin's invariant
scalar product from the Plancherel formula for the Whittaker
functions derived by Semenov-Tian-Shansky thus obtaining the natural
interpretation of the Sklyanin measure in terms of the Harish-Chandra
function.
\end{abstract}

\end{titlepage}
\clearpage \newpage

\setcounter{page}1
\footnotesize
\tableofcontents
\normalsize

\section{Introduction}
This paper is devoted to construction of the integral representation
for the eigenfunctions of the Hamiltonians for the quantum periodic Toda
chain. Our work was initiated by the recent results of Smirnov \cite{Sm}
concerning the structure of matrix elements in periodic Toda chain. The key
point of this approach is Sklyanin's invariant scalar product \cite{Skl1}
and
we tried to interpret it as a part of the Plancherel formula which relates
the scalar product of the eigenfunctions in original coordinate
representation
with those arising in the space of the separated variables. Such
interpretation forces to combine the analytic method of Gutzwiller \cite{Gu}
with the algebraic approach of Sklyanin \cite{Skl1}.

\noindent
To be more precise, Gutzwiller \cite{Gu} discovered an interesting
phenomenon which understood nowadays as quite universal one. If one tries
to find the eigenfunctions of the periodic Toda chain in terms of a formal
series in (auxiliary) eigenfunctions of the open chain with unknown
coefficients labelled by multiple index, then the multi-dimensional
difference equation on the coefficients thus obtained can be factorized to
the one-dimensional Baxter equations. This is what is called
the quantum separation of variables. In this way Gutzwiller obtained the
solution to the eigenvalue problem for the $N=2,3$ and 4-particle cases only
due to the lack of appropriate
 explicit solution for an arbitrary open Toda chain. The
next important step was done by Sklyanin \cite{Skl1}. He applied the
R-matrix
formalism to the present problem and simplified drastically the derivation
of
the Baxter equation to an arbitrary number of particles. The evident
advantages of Sklyanin's method are a wide area of application and effective
way to investigate different structures, for example, the invariant scalar
products. But, unfortunately, his purely algebraic method had no deal with
the analytical properties of auxiliary eigenfunctions of the specific model
and, therefore, some important information is inevitably lost to compare
with
the original Gutzwiller's approach. For example, as a price for efficiency
the measure is defined up to an arbitrary $i\hbar$-periodic function in
general and there is no natural way to fix this freedom uniquely;
there is a fundamental question what is the underlying principle
to determine the asymptotics and analytical properties of the solution to
Baxter equation and so on. Moreover, the transition from initial
coordinate-dependent wave function to those in separated variables
is somewhat obscured. The advantage of Gutzwiller approach consists of
explicit information concerning the analytical properties of auxiliary
eigenfunctions. As a consequence, it gives the possibility to calculate many
fine tuning quantities, for example the numerical coefficient in Baxter
equation, the asymptotics of its solutions etc. Therefore, there is a
natural
problem to generalize the Gutzwiller results to an arbitrary $N$-particle
Toda
chain and to extend them to other models. In particular, the investigation
of algebraic origin of the auxiliary functions arising in other
integrable models, is of higher importance.

\noindent
We consider the periodic Toda chain as a test example to future
generalization in these directions. This is the main motivation of the
present paper.

\noindent
We start from well-known observation that the Whittaker functions
\cite{Jac}-\cite{Ha} give the solution of the eigenvalue problem for the
general open Toda chain \cite{Konst, ST}. So it is naturally to use them to
construct the wave functions for the periodic Toda chain in the spirit of
Gutzwiller.
As the main result we obtain the integral representation
for the eigenfunctions of periodic Toda chain using intensively the
analytical properties of Whittaker functions investigated in
\cite{Jac}-\cite{Ha}. The second important counterpart of this
representation
is the explicit solution of the Baxter equation constructed by
Pasquier and Gaudin \cite{PG}.
We calculate the multiple integral by taking the residues thus
obtaining another representation for the wave function which
reproduces the particular results of Gutzwiller by putting $N=2, 3$ and
$N=4$
in our general formula. We observe also that the Plancherel formula obtained
by Semenov-Tian-Shansky \cite{ST} for the wavelets of Whittaker
functions
produce the natural way to calculate Sklyanin's scalar product. It turns out
that the Sklyanin measure is deeply connected with the Harish-Chandra
function arising in the theory of Whittaker functions.

\noindent
The paper is organized as follows.

\noindent
In Section 2 we briefly describe $N$-particle quantum periodic Toda chain in
the framework of $R$-matrix approach and formulate the spectral problem to
be
solved. The material here is quite standard.

\noindent
The solution of this problem is described by two theorems stated in Section
3.
This are the main results of the paper.

\noindent
The next four sections are devoted to the building blocks which allows to
prove the above theorems.

\noindent
The solution for the open $N\!-\!1$-particle Toda chain in
terms of Whittaker functions \cite{Konst,ST} is discussed in Section 4.
We introduce Weyl invariant Whittaker functions by simple renormalization of
the original ones and describe their properties using the classical
results \cite{Jac}-\cite{Ha}. We calculate also the scalar product of such
functions using the Plancherel formula derived in \cite{ST}.

\noindent
In Section 5 we follow the Gutzwiller approach \cite{Gu} to construct the
auxiliary functions which play the same role for the periodic Toda chain as
the exponentials for the standard Fourier transform. We generalize the
Gutzwiller solutions to an arbitrary $(N\!-\!1)$-particle solution in terms
of Weyl invariant Whittaker functions and calculate the action of the
diagonal operators of the monodromy matrix on the auxiliary
functions thus obtaining the functional relations which generalize the
corresponding relations for the Macdonald function.

\noindent
In Section 6 we reformulate the spectral problem using the (generalized)
Fourier transform to the so-called $\bgamma$-representation which is an
analytic version of the algebraic approach by Sklyanin \cite{Skl1}. The
analytical properties of the auxiliary function play a crucial role here. We
show that the wave function in $\bgamma$-representation is an entire
function
with definite asymptotics and it satisfies to multi-dimensional Baxter
equation. We derive the Plancherel
formula thus rigorously obtaining the scalar product in
$\bgamma$-variables introduced in \cite{Skl1}. As by-product, we obtain the
connection of the Harish-Chandra function \cite{Jac}-\cite{Ha} with
the Sklyanin measure.

\noindent
The solution to the Baxter equation is outlined in Section 7. We essentially
use the approach by Gutzwiller \cite{Gu} and especially those by Pasquier
and
Gaudin \cite{PG}.

\noindent
In Section 8 the proofs of the theorems stated in Section 3 are presented.

\section{Periodic Toda chain (description of the model)}
The quantum $N$-periodic Toda chain is a multi-dimensional eigenvalue
problem
with $N$ mutually commuting Hamiltonians
$H_k(x_0,p_0;\ldots;x_{{\N}-1},p_{{\N}-1})\,,\;(k=1,\ldots, N)$ where the
simplest Hamiltonians have the form
\be
H_1=\sum_{k=0}^{N-1}p_k\\
H_2=\sum_{k<m}p_kp_m-\sum_{k=0}^{N-1}e^{x_k-x_{k+1}}\\
H_3=\sum_{k<m<n}p_kp_mp_n+...
\ee
etc. and the phase variables $x_k,p_k$ satisfy to commutation relations
$[x_k,p_m]=i\hbar\delta_{km}$.
This system can be nicely described using the $R$-matrix approach
\cite{Skl1}.
It is well known that the Lax operator
\be\label{ln}
L_n(\la)\,=\, \left(\begin{array}{cc}\la-p_n & e^{-x_n}\\ -e^{x_n} & 0
\end{array}\right)
\ee
satisfies the commutation relations
\be
R(\la-\mu)(L_n(\la))\otimes I)(I\otimes L_n(\mu))=
(I\otimes L_n(\mu))(L_n(\la)\otimes I)R(\la-\mu)
\ee
where
\be
R(\la)=I\otimes I+\frac{i\hbar}{\la}\,P
\ee
is a rational $R$-matrix. The monodromy matrix
\be\label{a1}
T_{_N}(\la)\;\stackrel{\mbox{\tiny def}}{=}\;L_{N-1}(\la)\ldots
L_0(\la)\,\equiv\, \left(\begin{array}{cc}A(\la) & B(\la)\\
C(\la) & D(\la)\end{array}\right)
\ee
satisfies the analogous equation
\be\label{rtt}
R(\la-\mu)(T(\la)\otimes I)(I\otimes T(\mu))=
(I\otimes T(\mu))(T(\la)\otimes I)R(\la-\mu)
\ee
>From (\ref{rtt}) it can be easily shown that the trace of the monodromy
matrix
\be\label{tra}
\wh t(\la)=A(\la)+D(\la)
\ee
satisfies to commutation relations $[\wh t(\la),\wh t(\mu)]=0$ and
is a generating function for the Hamiltonians of the periodic Toda chain:
\be
\wh t(\la)=\sum_{k=0}^N(-1)^k\la^{N-k}H_k
\ee
Let us consider the spectral problem
\be
H_k\Psi_{\raise-3pt\hbox{$\scriptstyle\!\!\bfit E$}}\,=\,
E_k\Psi_{\raise-3pt\hbox{$\scriptstyle\!\!\bfit E$}}\hspace{1.5cm}
k=1,\ldots, N
\ee
where $\bfit E\equiv (E_1,\,\ldots\,,E_{\N})$. In other words,
$\Psi_{\raise-2pt\hbox{$\scriptstyle\!\!\!\bfit E $}}$
is an eigenfunction of the operator $\wh t(\la)$:
\be\label{sp3}
\wh t(\la)\Psi_{\raise-3pt\hbox{$\scriptstyle\!\!\bfit E $}}=t(\la;\bfit E)
\Psi_{\raise-3pt\hbox{$\scriptstyle\!\!\bfit E $}}
\ee
where
\be\label{eig}
t(\la;\bfit E)=\sum\limits_{k=0}^N(-1)^k\la^{N-k}E_k
\ee
For the future
convenience we shall denote the coordinate dependence of the solutions as
$\Psi_{\raise-2pt\hbox{$\scriptstyle\!\!\!\bfit E $}}(x_0,\bfit x)$
thus selecting $x_0$ from all other coordinates
$\bfit x=(x_1,\ldots,x_{{\N}-1})$. The reason for doing this will be clear
below. Evidently, the wave function has the following structure:
\be\label{kv0}
\Psi_{\raise-3pt\hbox{$\scriptstyle\!\!\bfit E $}}(x_0,\bfit x)=
\wt\Psi_{\raise-3pt\hbox{$\scriptstyle\!\!\bfit E $}}(x_0\!-\!x_1,\ldots,
x_{{\N}-2}\!-\!x_{{\N}-1})\,
\exp\left\{\frac{i}{\hbar}E_1\sum_{k=0}^{N-1}x_k\right\}
\ee
The main goal is to find the solution to (\ref{sp3}) such that
$\wt\Psi_{\raise-2pt\hbox{$\scriptstyle\!\!\!\bfit E $}}\in L^2(\RR^{N-1})$.
In equivalent terms, we impose the requirement
\be\label{kv}
\int f(E_1)\Psi_{\raise-3pt\hbox{$\scriptstyle\!\!\bfit E $}}(x_0,\bfit x)
dE_1\in L^2(\RR^N)
\ee
for any smooth function $f(y)\,,\;(y\in\RR)$ with finite support.

\section{Main results}
\begin{th}\label{th1}
The solution to the spectral problem (\ref{sp3}), (\ref{kv}) can be
represented as the integral over real variables
$\bgamma=(\gamma_1,\ldots,\gamma_{\N-1})$ in the following form:
\be\label{m1}
\Psi_{\raise-3pt\hbox{$\scriptstyle\!\!\bfit E $}}(x_0,\bfit x)=
\frac{(2\pi\hbar)^{-N}}{(N-1)!}\;\int\limits_{\r^{N-1}}\mu^{-1}
(\bgamma)C(\bgamma;\bfit E)\,\Psi_{\bgamma,E_1}(x_0,\bfit x)d\bgamma
\ee
where
\begin{itemize}
\item[(i)] The function $\mu(\bgamma)$ is defined by the formula
\be
\mu(\bgamma)=
\prod_{j<k}\left|\Gamma\Big(\frac{\gamma_j\!-\!\gamma_k}{i\hbar}\Big)\right|
^2
\ee
\item[(ii)] The function $C(\bgamma;\bfit E)$ is the solution of
multi-dimensional Baxter equation in the Pasquier-Gaudin form \cite{PG}
\be\label{pg1}
C(\bgamma;\bfit E)=\prod_{j=1}^{N-1}\frac{c_+(\gamma_j;\bfit E)-
\xi(\bfit E)c_-(\gamma_j;\bfit E)}
{\prod\limits_{k=1}^N\sinh\frac{\textstyle\pi}{\textstyle\hbar}
\Big(\gamma_j-\delta_k(\bfit E)\Big)}
\ee
where the entire functions $c_\pm(\gamma)$ are two Gutzwiller
solutions \cite{Gu} of one-dimensional Baxter equation
\be
t(\gamma;\bfit E)c(\gamma;\bfit E)=i^{-N}c(\gamma+i\hbar;\bfit E)+
i^{N}c(\gamma-i\hbar;\bfit E)
\ee
and the parameters $\xi(\bfit E),\;
\bfit\delta=(\delta_1(\bfit E),\ldots, \delta_{\N}(\bfit E)\,)$ satisfy the
Gutzwiller conditions (the energy quantization) \cite{Gu, PG}.
\item[(iii)] The wave function $\Psi_{\bgamma, E_1}(x_0,\bfit x)$ is defined
in terms of Whittaker function $w(\bfit x;\bgamma)$ \cite{Jac}-\cite{Ha}
for the $GL(N\!-\!1;\RR)$ open Toda chain \cite{Konst, ST}
\be\label{m6}
\!\!\Psi_{\bgamma, E_1}(x_0,\bfit x)=
\hbar^{-2i(\scriptstyle \bgamma,\brho)/\hbar}\prod_{j<k}\!\pi^{-1/2}
\Gamma\Big(\frac{\gamma_j\!-\!\gamma_k}{i\hbar}+\frac{1}{2}\Big)
\,w(\bfit x;\bgamma)\,
\exp\Big\{\frac{i}{\hbar}\Big(E_1\!-\!\sum_{k=1}^{N-1}\gamma_k\Big)x_0\Big\}
\ee
The function (\ref{m6}) is symmetric under the permutation of
$\gamma$-variables
\footnote{
In (\ref{m6}) we use the notation
$(\bgamma,\brho)\equiv\frac{1}{2}\sum\limits_{k=1}^{N-1}(N\!-\!2k)\gamma_k$.
}.
\end{itemize}
\end{th}

\bigskip\noindent
The multiple integral (\ref{m1}) can be explicitely evaluated. Let
$\bfit y=(y_1,\ldots,y_{\N})$ be an arbitrary vector in $\RR^N$. We denote
$\bfit y^{(s)}\equiv(y_1,\ldots,y_{s-1},y_{s+1},\ldots,y_{\N})\in\RR^{N-1}$
\begin{th}\label{th2}
Assuming that $\delta_j(\bfit E)\neq \delta_k(\bfit E)$,
the solution (\ref{m1}) can be written
(up to unessential numerical factor) in equivalent form
\be\label{gu0}
\Psi_{\raise-3pt\hbox{$\scriptstyle\!\!\bfit E $}}(x_0,\bfit x)=
\sum_{s=1}^N(-1)^{N-s}\!\!\sum_{\bfit n^{(s)}\in\z^{\N-1}}
\!\!\Delta(\bfit\delta^{(s)}\!+\!i\hbar\bfit n^{(s)})\,
C_+(\bfit\delta^{(s)}\!+\!i\hbar\bfit n^{(s)})\,
\Psi_{\bfit\delta^{(s)}+i\hbar\bfit n^{(s)},E_1}(x_0,\bfit x)
\ee
where
\be
C_+(\bgamma)\equiv\prod_{j=1}^{N-1}c_+(\gamma_j;\bfit E)
\ee
and $\Delta(\bgamma)=\prod\limits_{j>k}(\gamma_j-\gamma_k)$ is the
Vandermonde determinant.
\end{th}
\begin{rem}
For $N=2, 3$ and $N=4$ particle Toda chain the formula (\ref{gu0})
reproduces
the results obtained by Gutzwiller \cite{Gu}.
\end{rem}

\section{$GL(N\!-\!1,\RR)$ Toda chain and Whittaker function}
It can be easily shown that the operator
\be\label{au1}
B(\la)\,=\,e^{-x_0}\sum_{k=0}^{N-1}(-1)^k\la^{N-k-1}h_k(x_1,p_1;\ldots;
x_{\N-1},p_{\N-1})
\ee
is the generating function for the Hamiltonians $h_k$ of
$GL(N\!-\!1)$ Toda chain
\footnote{The commutativity of the operators $h_k$ follows
from the commutation relation $[B(\la),B(\mu)]=0$ which is encoded in
(\ref{rtt}).}.
In particular, the simplest Hamiltonians have the form
\be h_1=\sum_{k=1}^{N-1}p_k\\
h_2=\sum_{j<k}p_jp_k-\sum_{k=1}^{N-2}e^{x_k-x_{k+1}}
\ee
Let $\bgamma=(\gamma_1,\ldots,\gamma_{\N-1})$ be the set of
real parameters. We consider the spectral problem for $GL(N\!-\!1,\RR)$
Toda chain
\be\label{au3}
h_k\psi_\bgamma(\bfit x)=
\sigma_k(\bfit{\gamma})\psi_\bgamma(\bfit x)\hspace{1cm}k=1,\ldots, N-1
\ee
where $\sigma_k(\bgamma)$ are elementary symmetric functions. In equivalent
terms, $\psi_\bgamma(\bfit x)$ satisfies to the equation
\be\label{au4}
B(\la)\psi_\bgamma(\bfit x)=e^{-x_0}\prod_{m=1}^{N-1}(\la-\gamma_m)\,
\psi_\bgamma(\bfit x)
\ee
As a trivial corollary,
\be\label{au5}
B(\gamma_j)\psi_\bgamma(\bfit x)=0\hspace{2cm}\forall\,\gamma_j\in\bgamma
\ee
\begin{rem}
The equations (\ref{au5}) are an analog of the notion of the "operator
zeros"
introduced by Sklyanin \cite{Skl1}.
\end{rem}

\bigskip\noindent
Obviously, in the asymptotic region $x_{k+1}\gg x_k,\;(k=1,\ldots, N\!-\!2)$
all potentials vanish and the solution to (\ref{au3}) is a superposition
of plane waves. The main goal is to find the solution to (\ref{au3})
satisfying the following properties:
\begin{itemize}
\item[(i)] The solution vanishes very rapidly
\footnote{
More precisely,
$\psi_\bgamma(\bfit x)\sim\exp\{-\frac{2}{\hbar}e^{(x_k-x_{k+1})/2}\}$
as $x_k-x_{k+1}\to\infty$.
}
as $x_k-x_{k+1}\to\infty$ for any given $k$.
\item[(ii)] The function $\psi_\bgamma$ is Weyl-invariant, i.e. it is
symmetric under any permutation
\be\label{we}
\psi_{...\gamma_j...\gamma_k...}=\psi_{...\gamma_k...\gamma_j...}
\ee
\end{itemize}
The property (i) defines the unique solution (up to common
$\bgamma$-dependent factor). It is called the Whittaker function
$w(\bfit x;\bgamma)$ whose analytical and invariant properties were
studied in \cite{Jac}-\cite{Ha}.
In the context of the open Toda chain the Whittaker functions have been
appeared for the first time in \cite{Konst, ST}. The requirement of Weyl
invariance (\ref{we}) is very convenient for our purposes. It turns
out that symmetric function $\psi_\bgamma$ is most suitable to construct the
solution for the periodic Toda chain.

\bigskip\noindent
For $GL(N\!-\!1,\RR)$ Toda chain the Whittaker function has the following
integral representation \cite{Jac}-\cite{Ha}
\footnote{ The function (\ref{w3})
differs from the corresponding solution for the $SL(N\!-\!1,\RR)$ case
by the factor
$$
\exp\Big\{\frac{i}{(N\!-\!1)\hbar}\sum_{j=1}^{N-1}\gamma_j\cdot
\sum_{k=1}^{N-1}x_k\Big\}
$$
}
\be\label{w3}
w(\bfit x;\bgamma)=e^{\frac{i}{\hbar}(\bgamma,\bfit x)}
\int\limits_{-\infty}^\infty dz\prod_{k=1}^{N-2}
\Delta_k^{\frac{i}{\hbar}(\gamma_k-\gamma_{k+1})-\frac{1}{2}}(z)\,
\exp\Big\{\frac{2i}{\hbar}z_{k,k+1}e^{(x_k-x_{k+1})/2}\Big\}
\ee
where the integration is taken over the upper triangular
$(N\!-\!1)\!\times\!(N\!-\!1)$ matrix $z=||z_{jk}||$ with unit diagonal and
$\Delta_k(z)$ are the principal minors of the matrix $zz^t$.
Finally, $(.,.)$ is the standard scalar product in $\RR^{N-1}$.\\
In particular, taking $N=3$ and using the integral representation for the
Macdonald function $K_\nu(y)$, one obtains the appropriate solution
\be\label{ex}
w(\bfit x;\bgamma)=2\sqrt{\pi}\;
\frac{\phantom{aaa}\hbar^{-\frac{\gamma_1-\gamma_2}{\scriptstyle i\hbar}}}
{\Gamma\Big(\frac{\gamma_1-\gamma_2}{i\hbar}+\frac{1}{2}\Big)}\,
e^{\frac{i}{2\hbar}(\gamma_1+\gamma_2)(x_1+x_2)}
K_{\frac{\gamma_1-\gamma_2}{\scriptstyle i\hbar}}
\Big(\frac{2}{\hbar}e^{(x_1-x_2)/2}\Big)
\ee
of the spectral problem
\be
(p_1+p_2)w=(\gamma_1+\gamma_2)w\\
\left\{p_1p_2-e^{x_1-x_2}\right\}w=\gamma_1\gamma_2\,w
\ee
Indeed, $K_\nu(y)\sim y^{-1/2}e^{-y}$ as $y\to\infty$ and property (i) is
fulfilled.

\bigskip\noindent
All essential theorems for the classical Whittaker function
\cite{Jac}-\cite{Ha} hold in the present case as well.
We represent four lemmas below without proof
\footnote{
The notations and results are extracted from the papers \cite{Ha}.
}.
\begin{lem}\label{Weyl}
For any $s\in W$
\be\label{w4}
w(\bfit x;s\bgamma)={\cal M}(s;\bgamma)w(\bfit x;\bgamma)
\ee
where for the permutation $s_{jk}:\;\gamma_j\leftrightarrow\gamma_k\;(j<k)$
\be\label{w5}
{\cal M}(s_{jk};\bgamma)=
\prod_{m=j+1}^k\hbar^{2i(\gamma_k-\gamma_j)/\hbar}\;
\frac{\Gamma\Big(\frac{\gamma_j-\gamma_m}{i\hbar}+\frac{1}{2}\Big)
\Gamma\Big(\frac{\gamma_m-\gamma_k}{i\hbar}+\frac{1}{2}\Big)}
{\Gamma\Big(\frac{\gamma_m-\gamma_j}{i\hbar}+\frac{1}{2}\Big)
\Gamma\Big(\frac{\gamma_k-\gamma_m}{i\hbar}+\frac{1}{2}\Big)}
\ee
The coefficients ${\cal M}(s;\bgamma)$ satisfy to relations
\be\label{w6}
{\cal M}(s_1s_2;\bgamma)={\cal M}(s_1;s_2\bgamma){\cal M}(s_2;\bgamma)
\ee
\be\label{w7}
{\cal M}(s;\bgamma)\,\ov{\!\cal M}(s;\bgamma)=1
\hspace{1cm}{\rm Im}\,\gamma_k=0
\ee
(where $\ov z$ always denote the complex conjugation of $z$).\square
\end{lem}
\begin{lem}\label{ent}
The Whittaker function $w(\bfit x;\bgamma)$ can be analytically continued to
an entire function of $\bgamma\in\CC^{N-1}$.\square
\end{lem}
\begin{lem}
There is a unique solution $v(\bfit x;\bgamma)$ to (\ref{au3})
with the asymptotics
\be\label{Jost}
v(\bfit x;\bgamma)=e^{\frac{i}{\hbar}(\bgamma,\bfit x)}+
O\Big(\mbox{\rm max}\Big\{e^{x_k-x_{k+1}}\Big\}_{k=1}^{\N-2}\Big)
\ee
in the region $x_{k+1}\gg x_k,\;(k=1,\ldots,N-2)$.
Let $W$ be the Weyl group (the permutation group of the variables
$\bgamma$).
The functions $v(\bfit x;s\bgamma),\;(s\in W)$ form a basis of the
solutions to the spectral problem (\ref{au3}).\square
\end{lem}
\begin{lem}\label{bas1}
The Whittaker function (\ref{w3}) has the following expansion in terms of
basis functions $v(\bfit x;s\bgamma)\,,\;(s\in W)$:
\be\label{w1}
w(\bfit x;\bgamma)=\sum_{s\in W}c(s\bgamma){\cal M}^{-1}(s;\bgamma)\,
v(\bfit x;s\bgamma)
\ee
where
\be\label{w2}
c(\bgamma)=
\prod_{j<k}B\Big(\frac{\gamma_j\!-\!\gamma_k}{i\hbar}\,,\frac{1}{2}\Big)
\ee
is (non-normalized) Harish-Chandra function
\footnote{
In (\ref{w2}) $B(z,w)\equiv\frac{\textstyle\Gamma(z)\Gamma(w)}
{\textstyle\Gamma(z+w)}$.
}.\square
\end{lem}
Now we are able to define the Weyl symmetric solution and to describe
its properties.
Let $\brho$ is a half-sum of positive roots of $sl(N\!-\!1,\RR)$ written in
standard basis $\{\bfit e_k\}\in\RR^{N-1}$:
\be
\brho=\sum_{k=1}^{N-1}\rho_k\bfit e_k\,;\hspace{1cm}
\rho_k=\frac{1}{2}(N\!-\!2k)
\ee
\begin{lem}\label{psi1}
The function
\be\label{we2}
\psi_\bgamma(\bfit x)=\hbar^{-2i(\bgamma,\brho)/\hbar}
\prod_{j<k}\pi^{-1/2}\Gamma\Big(\frac{\gamma_j\!-\!\gamma_k}{i\hbar}\!+
\!\frac{1}{2}\Big)\;w(\bfit x;\bgamma)
\ee
\begin{itemize}
\item[(i)] is Weyl invariant.
\item[(ii)] it can be analytically continued
to an entire function of $\bgamma\in\CC^{N-1}$.
\item[(iii)] in terms of the basis $v(\bfit x;s\bgamma)\,,\;(s\in W)$
\be\label{exp2}
\psi_\bgamma(\bfit x)=\sum_{s\in W}N(s\bgamma)a(s\bgamma)\,
v(\bfit x;s\bgamma)
\ee
where $N(\bgamma)=\hbar^{-2i(\bgamma,\brho)/\hbar}$ and
\be\label{a}
a(\bgamma)=\prod_{j<k}\Gamma\Big(\frac{\gamma_j\!-\!\gamma_k}{i\hbar}\Big)
\ee
\item[(iv)] for real $\bgamma$ variables the function $\psi_\bgamma$ obeys
the asymptotics
\be\label{as0}
\psi_\bgamma(\bfit x)\sim |\gamma_k|^{\frac{2-N}{2}}
\exp\Big\{-\frac{\pi}{2\hbar}(N\!-\!2)|\gamma_k|\Big\}
\ee
as $|\gamma_k|\to\infty$.
\end{itemize}
\end{lem}
{\bf Proof.} The Weyl invariance is a direct corollary of the formulae
(\ref{w4}), (\ref{w5}).\\
To prove (ii), one should note that the poles coming from $\Gamma$ functions
in the r.h.s. of (\ref{we2}) are cancelled by the corresponding zeros of the
Whittaker function $w(\bfit x;\bgamma)$. Indeed, the function
${\cal M}(s_{jk};\bgamma)$ has the poles at
$\gamma_j-\gamma_k=-i\hbar(n+\frac{1}{2})\,,\;(n\in\NN)$. But the function
$w(\bfit x;s_{jk}\bgamma)$ is an entire function according to Lemma
\ref{ent}. Hence, from (\ref{w4}) it is clear that
$w(\bfit x;\bgamma)$ vanishes exactly at the same points.\\
To prove (iii), let
\be
b(\bgamma)\equiv
\prod_{j<k}\pi^{-1/2}\Gamma\Big(\frac{\gamma_j\!-\!\gamma_k}{i\hbar}\!+
\!\frac{1}{2}\Big)
\ee
for brevity. Then the Harish-Chandra function (\ref{w2}) is written
identically as a ratio
\be\label{ch2}
c(\bgamma)=\frac{a(\bgamma)}{b(\bgamma)}
\ee
while the definition (\ref{we2}) reads
\be\label{w8}
\psi_\bgamma(\bfit x)\equiv N(\bgamma)b(\bgamma)w(\bfit x;\bgamma)
\ee
Using the Weyl invariance $\psi_{s\bgamma}=\psi_\bgamma$,
one easily finds from (\ref{w4}) and (\ref{w8})
\be\label{w9}
{\cal M}(s;\bgamma)=\frac{N(\bgamma)}{N(s\bgamma)}\,
\frac{b(\bgamma)}{b(s\bgamma)}
\ee
Hence, the expansion (\ref{exp2}) is a simple corollary of (\ref{w1})
and definition (\ref{w8}).\\
The proof of (iv) follows from the Stirling formula for $\Gamma$-functions
and from the calculation of the integrals in (\ref{w3}) by the
stationary phase method
\footnote
{Actually, one can prove the exponential decreasing  $\psi_\bgamma\sim
\exp\Big\{\!\!-\textstyle\frac{\textstyle\pi}{\textstyle2\hbar}
(N\!-\!2)|\mbox{Re}\gamma_k|\Big\}$ as $|\mbox{Re}\,\gamma_k|\to\infty$
in any finite strip of the complex plane.}.
\square
\begin{lem}
Let $\bgamma=(\gamma_1,\ldots,\gamma_{\N-1})$ and
$\bgamma'=(\gamma'_1,\ldots,\gamma'_{\N-1})$be the real parameters. Then
\be\label{pl1}
\int\limits_{-\infty}^\infty\ov\psi_{\bgamma'}(\bfit x)
\psi_\bgamma(\bfit x)d\bfit x =(2\pi\hbar)^{N-1}
|a(\bgamma)|^2\sum_{s\in W}\delta(s\bgamma-\bgamma')
\ee
\end{lem}
{\bf Proof.} This is a consequence of the Plancherel formula
proved in \cite{ST} for the $SL(n,\RR)$ case. The scalar product (\ref{pl1})
results from Semenov-Tian-Shansky formula taking into account the relations
(\ref{w6}), (\ref{w7}) and (\ref{w9}).\square
\begin{con}
The Whittaker functions obey the completeness condition
\footnote{
We confirm this conjecture by explicit calculations for $GL(2,\RR)$ chain.
The general proof will be published elsewhere.
}
\be\label{con1}
\int\limits_{-\infty}^\infty |a(\bgamma)|^{-2}\psi_\bgamma(\bfit x)
\ov\psi_\bgamma(\bfit y)\,d\bgamma=(N\!-\!1)!\,(2\pi\hbar)^{N-1}
\delta(\bfit x-\bfit y)
\ee
\end{con}

\section{Auxiliary functions}
Let us introduce the {\it auxiliary} function
\be\label{gu1}
\Psi_{\bfit\gamma,\e}(x_0,\bfit x)\;\stackrel{\mbox{\tiny def}}{=}
\;e^{\raise3pt\hbox{$\frac{i}{\hbar}
\raise1pt\hbox{$\,\scriptstyle x_0 $}\scriptstyle\big(\e\,-
\!\sum\limits_{m=1}^{N-1}\gamma_m\big)$}}\,\psi_{\bfit\gamma}(\bfit x)
\ee
where $\e$ is an arbitrary real parameter. Obviously, this function
is the solutions of (\ref{au3}) or, in equivalent terms,
\be\label{bp2}
B(\la)\Psi_{\bfit\gamma,\e}(x_0,\bfit x)=
e^{-x_0}\prod_{m=1}^{N-1}(\la-\gamma_m)\,\Psi_{\bfit\gamma,\e}(x_0,\bfit x)
\ee
In particular,
\be\label{bp3}
B(\gamma_j)\Psi_{\bfit\gamma,\e}(x_0,\bfit x)=0\hspace{2cm}
\forall\,\gamma_j\in\bgamma
\ee
Further, by construction the function $\Psi_{\bgamma,\e}$ satisfies to
the equation
\be\label{me5}
H_1\Psi_{\bfit\gamma,\e}\,=\,\e\,\Psi_{\bfit\gamma, \e}
\ee
The scalar product of the auxiliary functions follows from (\ref{pl1}):
\be\label{pl2}
\int\limits_{-\infty}^\infty\ov\Psi_{\bfit\gamma',\e'}(x_0,\bfit x)
\Psi_{\bfit\gamma,\e}(x_0,\bfit x)dx_0d\bfit x=
(2\pi\hbar)^N\mu(\bgamma)\,
\delta(\e\!-\!\e')\sum_{s\in W}\delta(s\bgamma-\bgamma')
\ee
where
\be\label{pl3}
\mu(\bgamma)=\prod_{j<k}\left|\Gamma\Big(\frac{\gamma_j\!-\!\gamma_k}{i\hbar
}
\Big)\right|^2
\ee
Moreover, the auxiliary functions satisfy to completeness condition
\be\label{con2}
\int\limits_{-\infty}^\infty\mu^{-1}(\bgamma)\Psi_{\bgamma,\e}(x_0,\bfit x)
\ov\Psi_{\bgamma,\e}(y_0,\bfit y)\,d\bgamma\,d\e=
(N\!-\!1)!\,(2\pi\hbar)^N\delta(x_0-y_0)\delta(\bfit x-\bfit y)
\ee
provided the conjecture (\ref{con1}) holds.

\bigskip\noindent
Certainly, the function (\ref{gu1}) does not satisfy the whole system
(\ref{sp3}). Nevertheless, it possesses some remarkable properties which
allow to construct the eigenfunctions of the periodic Toda chain.

\bigskip\noindent
First of all, the functions $\Psi_{\bgamma,\e}$ and
$\ov\Psi_{\bgamma,\e}$ can be
extended to the entire functions of the complex variables
$\bgamma\!\in\!\CC^{N-1}$. This is an evident consequence of Lemma
\ref{psi1}.
We shall denote the corresponding analytic continuations by the same
letters.
Further, the following statement will be of importance below:
\begin{lem}
The action of the diagonal operators of the monodromy matrix (\ref{a1})
on auxiliary function $\Psi_{\bgamma,\e}\,\;(\bgamma\in\CC^{N-1})$
is defined by the formulae
\bse\label{adp}
\be\label{ap}
A(\gamma_j)\Psi_{\bgamma,\e}=i^N\Psi_{\bgamma+i\hbar\bfit e_j,\e}\\
\ee
\be\label{dp}
D(\gamma_j)\Psi_{\bgamma,\e}=i^{-N}\Psi_{\bgamma-i\hbar\bfit e_j,\e}
\ee
\ese
\end{lem}
{\bf Proof.}
We consider the following commutation relations encoded in (\ref{rtt}):
\be\label{db}
(\la-\mu+i\hbar)D(\mu)B(\la)\,=\,(\la-\mu)B(\la)D(\mu)+i\hbar D(\la)B(\mu)
\ee
\be\label{ab}
\!(\la-\mu+i\hbar)A(\la)B(\mu)\,=\,(\la-\mu)B(\mu)A(\la)+i\hbar A(\mu)B(\la)
\ee
Taking $\mu=\gamma_j$ in (\ref{db}) and applying this operator identity to
the
function $\psi_\bgamma(\bfit x)$, one arrives to the relation
\be\label{f7}
B(\la)D(\gamma_j)\psi_\bgamma(\bfit x)\,=\,e^{-x_0}\,(\la-\gamma_j+i\hbar)
\prod_{m\neq j}(\la-\gamma_m)\,D(\gamma_j)\psi_\bgamma(\bfit x)
\ee
since (\ref{au5}) holds
\footnote{Note that $D(\mu)$ does not contain the operator
$p_0$ and, therefore, commutes with $e^{-x_0}$.
}
.
Hence, $D(\gamma_j)\psi_\bgamma(\bfit x)$ is a solution to the spectral
problem (\ref{au4}) with the set of eigenvalues
\be\label{f8}
\bfit\gamma\;\to\;(\gamma_1,\ldots,\gamma_{j-1},\gamma_j-i\hbar,
\gamma_{j+1},\ldots,\gamma_{\N-1})
\ee
i.e. one obtains the expansion
\be\label{x0}
D(\gamma_j)\psi_\bgamma(\bfit x)=\sum_{s\in W}d_j(s;\bgamma)\,
v(\bfit x;s(\bgamma-i\hbar\bfit e_j))
\ee
due to completeness of the basis $v(\bfit x;\bgamma)\,\;(s\in W)$. On the
other hand, the function $D(\gamma_j)\psi_\bgamma(\bfit x)$ vanishes as
$x_k-x_{k+1}\to\infty$ for any given $k$. Hence, the r.h.s. in (\ref{x0}) is
proportional to $\psi_{\bgamma-i\hbar\bfit e_j}$ because of uniqueness of
the Whittaker function. Thus,
\be
D(\gamma_j)\psi_\bgamma(\bfit x)=d_j(\bgamma)\,
\psi_{\bgamma-i\hbar\bfit e_j}
\ee
To find the coefficients $d_j(\bgamma)$, we consider the asymptotics
in the region $x_{k+1}\gg x_k\,,\; (k=1,\ldots N-2)$
where (\ref{Jost}) holds. Note that in this limit
\be D(\la)=
-e^{x_{N-1}-x_0}\left\{\prod_{m=1}^{N-2}(\la-p_m)+\,
O\Big(\mbox{max}\Big\{e^{x_k-x_{k+1}}\Big\}_{k=1}^{\N-2}\Big)\right\}
\ee
Let $\bgamma$ be the complex variables with $\mbox{Im}\,\gamma_k\leq 0$.
Then it is easy to find that asymptotically
\be\label{x1}
D(\gamma_j)\psi_\bgamma(\bfit x)\sim
i^{-N}e^{-x_0+\frac{i}{\hbar}(\gamma_j-i\hbar)x_{N-1}}
\hbar^{-2i(\gamma_j-i\hbar)\rho_{N-1}/\hbar}
\prod_{m\neq j}\Gamma\Big(\frac{\gamma_m-\gamma_j+i\hbar}{i\hbar}\Big)\times
\\
\times
\wt\psi_{\gamma_1,\ldots,\gamma_{j-1}\gamma_{j+1},\ldots,\gamma_{N-1}}
(x_1,\ldots,x_{\N-2})
\ee
where $\wt\psi$ is the asymptotical value of $GL(N\!-\!2,\RR)$ Whittaker
function. On the other hand, under above conditions
\be\label{x2}
\psi_{\bgamma-i\hbar\bfit e_j}\sim
e^{\frac{i}{\hbar}(\gamma_j-i\hbar)x_{N-1}}
\hbar^{-2i(\gamma_j-i\hbar)\rho_{N-1}/\hbar}
\prod_{m\neq j}\Gamma\Big(\frac{\gamma_m-\gamma_j+i\hbar}{i\hbar}\Big)\times
\\
\times
\wt\psi_{\gamma_1,\ldots,\gamma_{j-1}\gamma_{j+1},\ldots,\gamma_{N-1}}
(x_1,\ldots,x_{\N-2})
\ee
The comparison of (\ref{x1}) and (\ref{x2}) gives
$d_j(\bgamma)=i^{-N}e^{-x_0}$. Hence, in the region
$\mbox{Im}\,\gamma_k\leq 0$
\be\label{dp2}
D(\gamma_j)\psi_\bgamma(\bfit x)=
i^{-N}e^{-x_0}\psi_{\bgamma-i\hbar\bfit e_j}(\bfit x)
\ee
and this relation can be obviously continued to the whole complex region.
The formula (\ref{dp}) follows then from the definition (\ref{gu1}).\\
Of course, it is possible to prove (\ref{ap}) according to the same
reasoning.
Actually, the relation (\ref{ap}) is a corollary of (\ref{dp}) since the
quantum determinant of the monodromy matrix (\ref{a1}) is equal to 1.
Applying the operator identity
\be
A(\gamma_j-i\hbar)D(\gamma_j)-C(\gamma_j-i\hbar)B(\gamma_j)=1
\ee
to the function $\Psi_{\bgamma,\e}$ and using the relations (\ref{bp3}) and
(\ref{dp}), one arrives to (\ref{ap}).  \square

\bigskip\noindent
As a direct consequence of (\ref{adp}) and definition (\ref{tra}),
for any $\bgamma\in\CC^{N-1}$
\be\label{tp}
\wh t(\gamma_j)\Psi_{\bgamma,\e}=i^N\Psi_{\bgamma+i\hbar\bfit e_j,\e}+
i^{-N}\Psi_{\bgamma-i\hbar\bfit e_j,\e}\\
\wh t^{\,\dagger}(\gamma_j)\ov\Psi_{\bgamma,\e}=
i^N\ov\Psi_{\bgamma+i\hbar\bfit e_j,\e}+
i^{-N}\ov\Psi_{\bgamma-i\hbar\bfit e_j,\e}\\
\ee
where $\wh t^{\,\dagger}(\la;x_0,p_0;\ldots;x_{\N-1},p_{\N-1})\equiv
\wh t(\la;x_0,-p_0;\ldots;x_{\N-1},-p_{\N-1})\,,\;(\la\in\CC)$
and the analytic continuation of the functions is understood.
\begin{rem}
The formula (\ref{tp}) has been derived by Gutzwiller \cite{Gu} using
cumbersome explicit calculations for $n=2$ and $n=3$-particle open Toda
chain.
\end{rem}
To prove Theorem \ref{th1} one needs to determine the action of the
operator $\wh t(\la)\;(\la\in\CC)$ on the function $\Psi_{\bfit\gamma,\e}$.
\begin{lem}
\be\label{tl}
\wh t(\la)\Psi_{\bfit\gamma,\e}=\\ =
\Big(\la\!-\!\e\!+\!\sum_{j=1}^{N-1}\!\gamma_j\Big)
\prod_{m=1}^{N-1}(\la\!-\!\gamma_m)\,
\Psi_{\bfit\gamma,\e}\!+\!\sum_{j=1}^{N-1}\prod_{m\neq j}
\frac{\la\!-\!\gamma_m}{\gamma_j\!-\!\gamma_m}
\Big(i^N\Psi_{\bfit\gamma+i\hbar\bfit e_j,\e}+
i^{-N}\Psi_{\bfit\gamma-i\hbar\bfit e_j,\e}\Big)
\ee
\end{lem}
{\bf Proof.} The function $\wh t(\la)\Psi_{\bfit\gamma,\e}$ is a polynomial
in $\lambda$ of order $N$ with the leading terms
\be
\wh t(\la)\Psi_{\bfit\gamma,\e}=(\la^N-\la^{N-1}\e)\Psi_{\bfit\gamma,\e}+
O(\la^{N-2})
\ee
according to (\ref{me5}). Moreover, this polynomial is defined at the points
$\la=\gamma_j\,,\;(j=1,\ldots, N-1)$ due to (\ref{tp}). Using the
standard interpolation formula, one easily obtains (\ref{tl}). \square
\begin{rem}
The formula (\ref{tl}) is a coordinate version of the operator interpolation
formula obtained in \cite{Skl1}.
\end{rem}

\section{Periodic Toda chain in $\bgamma$-representation}
Let $\Psi_{\raise-2pt\hbox{$\scriptstyle\!\!\!\bfit E $}}(x_0,\bfit x)$
be the fast decreasing solution of the problem (\ref{sp3}).
We define the function $C(\bgamma)$ by generalized Fourier transform:
\be\label{c1}
\delta(E_1-\e)\,C(\bgamma;\bfit E)=\int\limits_{-\infty}^\infty
\Psi_{\raise-3pt\hbox{$\scriptstyle\!\!\bfit E $}}(x_0,\bfit x)
\,\ov\Psi_{\bfit\gamma,\e}(x_0,\bfit x)dx_0d\bfit x
\ee
\begin{th}\label{thc}
The function $C(\bgamma)$ possesses the following properties:
\begin{itemize}
\item[(i)] It is a symmetric function with respect to $\bgamma$-variables.
\item[(ii)] It is an entire function of the variables $\bgamma\in\CC^{N-1}$.
\item[(iii)] The function $C(\bgamma)$ obeys the asymptotics
\be\label{as1}
C(\bgamma)\sim |\gamma_k|^{-N/2}e^{-\frac{\pi N|\gamma_k|}{2\hbar}}
\ee
as $|\gamma_k|\to\infty$ along the real axes.
\item[(iv)] The function $C(\bgamma)$ satisfies the multi-dimensional Baxter
equation
\be\label{b1}
t(\gamma_j;\bfit E)C(\bgamma;\bfit E)=i^NC(\bgamma+i\hbar\bfit e_j;\bfit E)
+i^{-N}C(\bgamma-i\hbar\bfit e_j;\bfit E)
\ee
where $t(\gamma;\bfit E)$ is defined by (\ref{eig}).
\end{itemize}
\end{th}
{\bf Proof.}
The symmetricity of the function $C(\bgamma)$ is obvious.\\
We present here only the sketch of the proof concerning the statements
(ii) and (iii).\\
The statement (ii) follows from the assertion that the auxiliary function
$\Psi_{\bgamma,\e}$ is an entire one while the solution of the
periodic chain vanishes very rapidly as $|x_k\!-\!x_{k+1}|\to\infty$.
\footnote{
Actually, the boundary conditions have the same importance here as the
requirement of compact support in the theory of the analytic
continuation for the usual Fourier transform.
}.\\
(iii) The asymptotics (\ref{as1}) is a combination of two factors.
The first one comes from the asymptotics (\ref{as0}) which results from the
Stirling formula for the $\Gamma$-functions in the definition (\ref{we2}).
The second factor $\sim |\gamma_k|^{-1}\exp\{-\pi|\gamma_k|/\hbar\}$ results
from the stationary phase method while calculating the multiple integral
including the Whittaker function (\ref{w3}). The calculation is based
heavily
on the exact asymptotics of the function
$\Psi_{\raise-2pt\hbox{$\scriptstyle\!\!\!\bfit E $}}(x_0,\bfit x)$ as
$|x_k-x_{k+1}|\to\infty$.\\
The proof of (iv) is simple. Using the definition (\ref{sp3}) and
integrating by parts (evidently, boundary terms vanish), one obtains
\be
\delta(E_1-\e)\,t(\gamma_j;\bfit E)C(\bgamma)\equiv
\int\limits_{-\infty}^\infty
\Big\{\wh t(\gamma_j)\Psi_{\raise-3pt\hbox{$\scriptstyle\!\!\bfit E
$}}(x_0,\bfit x)\Big\}
\,\ov\Psi_{\bfit\gamma,\e}(x_0,\bfit x)dx_0d\bfit x =\\
=\int\limits_{-\infty}^\infty
\Psi_{\raise-3pt\hbox{$\scriptstyle\!\!\bfit E $}}(x_0,\bfit x)
\,\Big\{\wh t^{\,\dagger}(\gamma_j)\ov\Psi_{\bfit\gamma,\e}(x_0,\bfit x)
\Big\}dx_0d\bfit x
\ee
Taking into account the relation (\ref{tp}), the Baxter equation (\ref{b1})
follows from definition (\ref{c1}).\square

\bigskip\noindent
Let us consider the following function
\footnote{The expression (\ref{in1}) is an inversion formula to
(\ref{c1}) {\it assuming} the completeness condition (\ref{con2}). Otherwise
one can only conclude that (\ref{c1}) is a corollary of (\ref{in1})
due to the scalar product (\ref{pl2}).
}
\be\label{in1}
\Psi_{\raise-3pt\hbox{$\scriptstyle\!\!\bfit E $}}(x_0,\bfit x)=
\frac{(2\pi\hbar)^{-N}}{(N\!-\!1)!}
\int\limits_{-\infty}^\infty\mu^{-1}(\bgamma)\,
C(\bgamma;\bfit E)\,
\Psi_{\bfit\gamma,E_1}(x_0,\bfit x)d\bgamma
\ee
The integral (\ref{in1}) integral is correctly defined. Indeed, the measure
\be\label{mu1}
\mu^{-1}(\bgamma)=
\prod_{j<k}\left|\Gamma\Big(\frac{\gamma_j\!-\!\gamma_k}{i\hbar}
\Big)\right|^{-2}=\\ =
(\pi\hbar)^{-(\N-1)(\N-2)/2}\prod_{j<k}(\gamma_j\!-\!\gamma_k)
\sinh\frac{\pi}{\hbar}(\gamma_j\!-\!\gamma_k)
\ee
is an entire function. Therefore, there are no poles in the integrand.
Moreover,
\be\label{mu2}
\mu^{-1}(\bgamma)\sim|\gamma_k|^{N-2}
\exp\Big\{\frac{\pi}{\hbar}(N\!-\!2)|\gamma_k|\Big\}
\ee
as $|\gamma_k|\to\infty$. Taking into account the asymptotics (\ref{as0})
and
(\ref{as1}) one concludes that the integrand has the behavior
$\sim|\gamma_k|^{-1}\exp\{-\pi|\gamma_k|/\hbar\}$ as
$|\gamma_k|\to\infty$. Therefore, the integral (\ref{in1}) is convergent.

\bigskip\noindent
Using the scalar product (\ref{pl2}), one can write the
Plancherel formula
\footnote{
Let us stress again that (\ref{top}) is deeply connected with the Plancherel
formula for $SL(n,\RR)$ Toda chain derived in \cite{ST}.
}
\be\label{top}
(2\pi\hbar)^N\!\int\limits_{-\infty}^\infty
\ov\Psi_{\raise-3pt\hbox{$\scriptstyle\!\!\bfit E'$}}(x_0,\bfit x)
\Psi_{\raise-3pt\hbox{$\scriptstyle\!\!\bfit E $}}(x_0,\bfit x)dx_0d\bfit x
=\frac{1}{(\!N\!\!-\!\!1\!)!}\,\delta(E_1\!-\!E_1')\!
\int\limits_{-\infty}^\infty\!\mu^{-1}(\bgamma)\,\ov C(\bgamma;\bfit E')
C(\bgamma;\bfit E)d\bgamma
\ee
The integral in the r.h.s. of (\ref{top}) is absolutely convergent due to
asymptotics (\ref{as1}) and (\ref{mu2}). Hence, the norm
$||\Psi_{\raise-2pt\hbox{$\scriptstyle\!\!\!\bfit E $}}||$
is finite modulo $GL(1)$ $\delta$-function $\delta(E_1\!-\!E_1')$
(see corresponding factor in (\ref{kv0}) which results to this function).
\begin{rem}
The scalar product in $\bgamma$-variables appears for the first time in
\cite{Skl1}. We see that the Sklyanin measure $\mu^{-1}(\bgamma)$ is
naturally connected with the Harish-Chandra function (\ref{ch2}).
\end{rem}

\bigskip\noindent
To prove that the function (\ref{in1}) satisfies to the spectral problem
(\ref{sp3}), one needs more details information concerning the asymptotics
of function $C(\bgamma;\bfit E)$: it should
exponentially decrease in any finite strip of the complex plane
$\gamma_k\in\CC$ as $|\mbox{\rm Re}\,\gamma_k|\to\infty$. Then (\ref{in1})
is
a solution if $C(\bgamma;\bfit E)$ satisfies the Baxter equation (\ref{b1})
(see the proof of Theorem \ref{th1} below). Up to now we can prove only the
asymptotics (\ref{as1}) starting from the definition (\ref{c1}) (though the
analytic continuation seems almost evident). Therefore, in the next section
we describe the explicit solution to equation (\ref{b1}) \cite{Gu, PG} which
obviously possesses the required asymptotics.

\section{Solution of Baxter equation}
In this section the Pasquier-Gaudin solution \cite{PG} to the Baxter
equation (\ref{b1}) is described which leads to refined derivation of
Gutzwiller quantization condition \cite{Gu}.

\bigskip\noindent
We solve the equation (\ref{b1}) in the separated form
\be\label{ss}
C(\bgamma;\bfit E)=\prod_{j=1}^{N-1}c(\gamma_j;\bfit E)
\ee
Obviously, the "one particle" functions $c(\gamma)$ satisfy to
the one-dimensional Baxter equation
\be\label{fb1}
t(\gamma;\bfit E)\,c(\gamma;\bfit E)=
i^{N}c(\gamma+i\hbar;\bfit E)+i^{-N}c(\gamma-i\hbar;\bfit E)
\ee
where $t(\gamma;\bfit E)$ is, by definition, a polynomial
$\prod\limits_{k=1}^N(\gamma-\la_k(\bfit E))$.
We assume that the roots $\la_k(\bfit E)$ are real for the solutions with
the finite norm
\footnote{This can be demonstrated directly for $N=2,3$ particle chain.
}.
We shall oftenly do not write the explicit dependence
on energies for brevity.
\begin{lem}
\cite{Gu} The equation (\ref{fb1}) admits two fundamental solutions
\be\label{fs0}
\wt c_\pm(\gamma)=
e^{\raise2pt\hbox{${\scriptstyle-}\frac{\pi N\gamma}{\hbar}$}}c_\pm(\gamma)
\ee
where
\be\label{fs}
c_\pm(\gamma)\,=\,\frac{K_\pm(\gamma)}
{\prod\limits_{k=1}^N
\hbar^{\!\raise2pt\hbox{${\scriptstyle \mp}\frac{i\gamma}{\hbar}$}}
\Gamma\Big(1\mp\frac{\textstyle i}{\textstyle \hbar}(\gamma-\la_k)\Big)}
\ee
and $K_\pm(\gamma)$ be the following $\NN\times\NN$ determinants:
\bse\label{k1}
\be\label{k1a}
K_+(\gamma)\equiv
\left|
\begin{array}{cccccc}
1 &\frac{1}{t(\gamma\!+\!i\hbar)} & 0 & \ldots & \ldots & \ldots
\\
\frac{1}{t(\gamma\!+\!2i\hbar)} &1 & \frac{1}{t(\gamma\!+\!2i\hbar)}
& 0 & \ldots & \ldots  
\\
0 &\frac{1}{t(\gamma\!+\!3i\hbar)} &1
&\frac{1}{t(\gamma\!+\!3i\hbar)}& 0  & \ldots  
\\
\ldots &\ldots &\ldots & \ldots & \ldots & \ldots  
\end{array}
\right|
\ee
\be\label{k1b}
K_-(\gamma)\,\equiv\,
\left|
\begin{array}{cccccc}
\ldots  & \ldots & \ldots & \ldots & \ldots & \ldots\\
\ldots & 0& \frac{1}{t(\gamma\!-\!3i\hbar)}& 1 &
\frac{1}{t(\gamma\!-\!3i\hbar)} & 0\\
\ldots &\ldots  & 0 &\frac{1}{t(\gamma\!-\!2i\hbar)} & 1  &
\frac{1}{t(\gamma\!-\!2i\hbar)}\\
\ldots &\ldots & \ldots & 0 & \frac{1}{t(\gamma\!-\!i\hbar)} & 1
\end{array}
\right|
\ee
\ese
\end{lem}
{\bf Proof.} The determinants $K_\pm(\gamma)$ are correctly defined
\cite{WW} and satisfy to the equations
\be\label{ab1a}
K_+(\gamma-i\hbar)\,=\,K_+(\gamma)-
\frac{1}{t(\gamma)t(\gamma\!+\!i\hbar)}K_+(\gamma+i\hbar)\\
K_-(\gamma+i\hbar)\,=\,K_-(\gamma)-
\frac{1}{t(\gamma)t(\gamma\!-\!i\hbar)}K_-(\gamma-i\hbar)
\ee
This can be easily proved by expansion of these determinant with respect to
the first row. The rest is trivial. Note that solutions (\ref{fs})
are determined up to arbitrary $i\hbar$-periodic function.\square

\vspace{-0.5cm}
\begin{rem}
The function $c_\pm(\gamma)$ are the solutions of the Baxter equation
\be\label{fb2}
t(\gamma)\,c(\gamma)\,=\,i^{-N}c(\gamma+i\hbar)\,+\,i^N c(\gamma-i\hbar)
\ee
\end{rem}

\vspace{-0.5cm}
\begin{lem}\label{las}
The functions $c_\pm(\gamma)$ possess the following properties:
\begin{itemize}
\item[(i)]
$c_\pm(\gamma)$ are the entire functions in $\gamma$.
\item[(ii)] Let $\gamma$ be real. Then
\be\label{as5}
c_\pm(\gamma)\sim\left\{
\begin{array}{ll}
|\gamma|^{-N/2}\;
e^{\raise2pt\hbox{$\frac{\pi N|\gamma|}{2\hbar}$}}\,
\exp\left\{\pm\frac{iN\gamma}{\hbar}\log\frac{|\gamma|}{e}\pm
\frac{\pi iN}{4}\right\}\hspace{1cm} & \gamma\to+\infty\\
|\gamma|^{-N/2}\;
e^{\raise2pt\hbox{$\frac{\pi N|\gamma|}{2\hbar}$}}\,
\exp\left\{\pm\frac{iN\gamma}{\hbar}\log\frac{|\gamma|}{e}
\mp\frac{\pi iN}{4}\right\}\hspace{1cm}  & \gamma\to -\infty
\end{array}\right.
\ee
\item[(iii)]
\be\label{as6}
c_+(\gamma)\sim \exp\Big\{\!-\frac{N|\gamma|}{\hbar}\log|\gamma|\Big\}
\hspace{2cm}\gamma\to+i\infty\\
c_-(\gamma)\sim \exp\Big\{\!-\frac{N|\gamma|}{\hbar}\log|\gamma|\Big\}
\hspace{2cm}\gamma\to-i\infty
\ee
\end{itemize}
\end{lem}
{\bf Proof.}
(i) It is easy to see that the poles of $K_\pm(\gamma)$
are cancelled by the corresponding poles coming from the product of
$\Gamma$-functions in (\ref{fs}). Hence, $c_\pm(\gamma)$ are an entire
functions.\\
(ii) It is clear that
$\lim\limits_{|\gamma|\to\infty}K_\pm(\gamma)=1$ for real $\gamma$.
Similarly, $K_\pm(\gamma)\to 1$ as $\gamma\to\pm i\infty$.
The asymptotics follow from the Stirling formula for $\Gamma$-functions.
Note that the Stirling formula gives even more detailed asymptotics
than (\ref{as5}); actually
$c_\pm(\gamma)\sim \exp\{\frac{\pi N}{2\hbar}|\mbox{Re}\,\gamma|\}$ as
$|\mbox{Re}\,\gamma|\to\infty$ in any finite strip
$|\mbox{Im}\,\gamma|\leq\e,\;(\e\geq 0)$.
\square

\bigskip\noindent
In order to construct the solution with the asymptotics (\ref{as1}), one
should recall that the solutions (\ref{fs0}) are defined up to any
$i\hbar$-periodic factor. Clearly, the factor
\be\label{fact}
\prod_{k=1}^N\frac{e^{\raise2pt\hbox{$\frac{\pi\gamma}{\hbar}$}}}
{\sinh\frac{\textstyle\pi}{\textstyle\hbar}(\gamma\!-\!\delta_k)}
\ee
(where $\delta_k$ are an arbitrary parameters)
does not spoil the asymptotics of the functions (\ref{fs}) as
$\gamma\to+\infty$ along the real axis while giving the correcting factor
$e^{\raise3pt\hbox{${\scriptstyle-}\frac{2\pi|\gamma|}{\hbar}$}}$ as
$\gamma\to-\infty$. Hence, one can consider the following solution of the
Baxter equation (\ref{fb1}):
\be\label{true1}
c(\gamma)=
\frac{c_+(\gamma)-\xi c_-(\gamma)}
{\prod\limits_{k=1}^{\N}\sinh
\frac{\textstyle\pi}{\textstyle\hbar}(\gamma\!-\!\delta_k)}
\ee
where $\xi$ is a arbitrary constant. Due to (\ref{as5}) this solution has
prescribed asymptotics
\be
c(\gamma)\sim |\gamma|^{-N/2}
e^{\raise2pt\hbox{${\scriptstyle-}\frac{\pi N|\gamma|}{2\hbar}$}}
\ee
as $\gamma\to\pm\infty$.
On the other hand, the denominator in (\ref{true1})
has the infinite number of poles at $\gamma=\delta_k+i\hbar
n_k\,,\;n_k\in\ZZ,\,k=1,\ldots, N$ and the solution (\ref{true1}) is not an
entire function in general. The poles are cancelled only if the following
conditions hold:
\be\label{q1}
c_+(\delta_k+i\hbar n_k)=\xi c_-(\delta_k+i\hbar n_k)
\ee
In turn, this means that the Wronskian
\be\label{wr1}
W(\gamma)=c_+(\gamma)c_-(\gamma+i\hbar)-
c_+(\gamma+i\hbar)c_-(\gamma)
\ee
vanishes at $\gamma=\delta_k+i\hbar n_k$.
Due to Baxter equation (\ref{fb2})
\be\label{wr2}
W(\gamma+i\hbar)=(-1)^NW(\gamma)
\ee
Substitution of (\ref{fs}) to (\ref{wr1}) gives
\be\label{wr3}
W(\gamma)=i^{-N}\ch(\gamma)
\,\prod_{k=1}^N\pi^{-1}\hbar\,\sinh\frac{\pi}{\hbar}(\gamma-\la_k)
\ee
where the function
\be\label{wr4}
\ch(\gamma)\equiv K_+(\gamma)K_-(\gamma+i\hbar)-
\frac{K_+(\gamma+i\hbar)K_-(\gamma)}{t(\gamma)t(\gamma+i\hbar)}
\ee
is called the Hill determinant \cite{WW}.
\begin{lem}\cite{WW},\cite{Gu} The function $\ch(\gamma)$ has the following
three-diagonal determinant form:
\be\label{hill}
\!\ch(\gamma)=
\left|
\begin{array}{cccccccc}
\ldots  &\ldots & \ldots & \ldots & \ldots & \ldots & \ldots\\
\ldots  &1 & \frac{1}{t(\gamma\!-\!i\hbar)}
& 0  & \ldots & \ldots & \ldots\\
\ldots &\frac{1}{t(\gamma)}& 1& \frac{1}{t(\gamma)} & 0 & \ldots & \ldots\\
\ldots &0 &\frac{1}{t(\gamma\!+\!i\hbar)}& 1 &
\frac{1}{t(\gamma\!+\!i\hbar)}
& 0 & \ldots\\
\ldots & \ldots & \ldots & \ldots & \ldots & \ldots & \ldots\\
\end{array}
\right|
\ee
This infinite determinant is absolutely convergent;
$\ch(\gamma)$ is a meromorphic function with the poles at the points
$\la_k+in_k\hbar\,,\;(n_k\in\ZZ,\,k=1,\ldots,N)$.
The Hill determinant is $i\hbar$-periodic
functions with the asymptotics $\ch(\gamma)=1+O(\gamma^{-1})$ as
$\gamma\to\pm\infty$.
The following representation is valid:
\be\label{q6}
\ch(\gamma)\,=\,1+\sum_{k=1}^N\al_k(\bfit E)\,
\coth\frac{\pi}{\hbar}\Big(\gamma-\la_k(\bfit E)\Big)
\ee
where $\al_k(\bfit E)$ are some definite functions depending on the
parameters
$\bfit E$ only. These functions satisfy to the constraint
\be\label{q4}
\sum_{k=1}^N\al_k(\bfit E)\,=\,0
\ee
The equation $\ch(\gamma)=0$ has exactly $N$ real solutions
$\delta_k(\bfit E),\,(k=1,\ldots,N) $ subjected to the constraint
\be
\sum_{k=1}^N\delta_k(\bfit E)=\sum_{k=1}^N\la_k(\bfit E)
\ee\square
\end{lem}
Due to its periodicity,  $\ch(\gamma)=0$ at any point
$\delta_k(\bfit E)+i\hbar n_k\;(n_k\in\ZZ)$. Hence, the solution
(\ref{true1}) has no poles if one takes $\delta_k=\delta_k(\bfit E)$
provided
that the constant $\xi$ is chosen in such a way that
\be\label{xi}
\hspace{2cm}
\xi=\left.\frac{c_+(\gamma)}{c_-(\gamma)}\right|_{\gamma=\delta_k(\bfit E)}
\hspace{2cm}k=1,\ldots,N
\ee
Hence, one arrives to the following
\begin{lem}\cite{PG}
The function
\be\label{pg2}
C(\bgamma;\bfit E)=\prod_{j=1}^{N-1}\frac{c_+(\gamma_j;\bfit E)-
\xi(\bfit E)c_-(\gamma_j;\bfit E)}
{\prod\limits_{k=1}^N\sinh\frac{\textstyle\pi}{\textstyle\hbar}
\Big(\gamma_j-\delta_k(\bfit E)\Big)}
\ee
where $\delta_k(\bfit E)$ are the zeros of the Hill determinant (\ref{hill})
and the constant $\xi$ is chosen according to (\ref{xi}), satisfies to
conditions of Theorem \ref{thc}.
\end{lem}
The quantization conditions
\be\label{q2}
\frac{c_+(\delta_1)}{c_-(\delta_1)}=\ldots=
\frac{c_+(\delta_{\N})}{c_-(\delta_{\N})}
\ee
determine the energy spectrum of the problem. These have been obtained for
the first time by Gutzwiller \cite{Gu} using quite different method.

\section{Proof of the main theorems}\label{prfs}
{\bf Proof of Theorem \ref{th1}.} We have proved already that the function
$\Psi_{\raise-2pt\hbox{$\scriptstyle\!\!\!\bfit E $}}$ is integrable in
a sense of (\ref{top}), (\ref{kv}). Now we prove that the
function (\ref{in1}) satisfies to the spectral problem (\ref{sp3}) if the
entire solution $C(\bgamma;\bfit E)$ to the Baxter equation (\ref{b1})
obeys the asymptotics $\sim\exp\{-\frac{\pi
N}{2\hbar}|\mbox{Re}\,\gamma_j|\}$
as $|\mbox{Re}\,\gamma_j|\to\infty$ in the strip
$-i\hbar\leq\mbox{Im}\,\gamma_j\leq i\hbar$. Note that the
Pasquier-Gaudin solution (\ref{pg2}) satisfies to this requirement (see
proof
of
Lemma \ref{las}).

\bigskip\noindent
Due to (\ref{tl}) one obtains
\footnote{
We discard unessential  numerical factor in (\ref{in1}).
}
\be\label{start}
\wh t(\la)\Psi_{\raise-3pt\hbox{$\scriptstyle\!\!\bfit E $}}=
\int\limits_{-\infty}^\infty d\bgamma\,\mu^{-1}(\bgamma)\,
C(\bgamma)\times\\
\left\{\Big(\la\!-\!E_1\!+\!\sum_{j=1}^{N-1}\!\gamma_j\Big)
\prod_{m=1}^{N-1}(\la\!-\!\gamma_m)\,
\Psi_{\bfit\gamma,E_1}\!+\!\sum_{j=1}^{N-1}\prod_{m\neq j}
\frac{\la\!-\!\gamma_m}{\gamma_j\!-\!\gamma_m}
\Big(i^N\Psi_{\bfit\gamma+i\hbar\bfit e_j,E_1}+
i^{-N}\Psi_{\bfit\gamma-i\hbar\bfit e_j,E_1}\Big)\right\}
\ee
Changing $\gamma_j\to \gamma_j\pm i\hbar$ in appropriate parts of the
integrand and noting that the measure (\ref{mu1}) satisfies to the
difference
equation
\be\label{pl4}
\mu^{-1}(\bfit\gamma+i\hbar\bfit\delta_j)\,=\,(-1)^N\mu^{-1}(\bfit\gamma)\,
\prod_{m\neq j}\frac{\gamma_j-\gamma_m\!+\!i\hbar}{\gamma_j-\gamma_m}
\ee
one arrives to the expression
\be\label{zz}
\wh t(\la)\Psi_{\raise-3pt\hbox{$\scriptstyle\!\!\bfit E $}}=
\int\limits_{-\infty}^\infty d\bgamma\,\mu^{-1}(\bgamma)\,
C(\bgamma)\Big(\la\!-\!E_1\!+\!\sum_{j=1}^{N-1}\!\gamma_j\Big)
\prod_{m=1}^{N-1}(\la\!-\!\gamma_m)\,
\Psi_{\bfit\gamma,E_1}+\\
+i^N\sum_{j=1}^{N-1}\int\limits_{-\infty}^\infty d\gamma_1\ldots\int\limits
_{-\infty-i\hbar}^{\infty-i\hbar}d\gamma_j\ldots\int\limits_{-\infty}^\infty
d\gamma_{{\N}-1}\,\mu^{-1}(\bgamma)
C(\bgamma\!+\!i\hbar\bfit e_j)\Psi_{\bgamma,E_1}\,
\prod_{m\neq j} \frac{\la\!-\!\gamma_m}{\gamma_j\!-\!\gamma_m} + \\ +
i^{-N}\sum_{j=1}^{N-1}\int\limits_{-\infty}^\infty d\gamma_1\ldots
\int\limits_{-\infty+i\hbar}^{\infty+i\hbar}d\gamma_j\ldots
\int\limits_{-\infty}^\infty d\gamma_{{\N}-1}\,\mu^{-1}(\bgamma)
C(\bgamma\!-\!i\hbar\bfit e_j)\Psi_{\bgamma,E_1}\,
\prod_{m\neq j} \frac{\la\!-\!\gamma_m}{\gamma_j\!-\!\gamma_m}
\ee
The integration contours $[-\infty\pm i\hbar,\infty\pm i\hbar]$ can be
deformed to the standard one $[-\infty,\infty]$. Indeed, the corresponding
integrands are the entire functions (all possible poles are cancelled by
appropriate zeros of the measure $\mu^{-1}(\bgamma)$, see (\ref{mu1})).
Moreover, it is not hard to see that the integrand
is exponentially decreasing in the strip
$-i\hbar\leq\mbox{Im}\,\gamma_j\leq i\hbar$
as $|\mbox{Re}\,\gamma_j|\to\infty$. Therefore, the contour integral around
the strip vanishes and all integrals in (\ref{zz}) can be deformed to the
standard ones. Finally, using the Baxter equation (\ref{b1}), one
obtains from (\ref{zz})
\be\label{zz2}
\wh t(\la)\Psi_{\raise-3pt\hbox{$\scriptstyle\!\!\bfit E $}}=\\ =
\int\limits_{-\infty}^\infty d\bgamma\,\mu^{-1}(\bgamma)\,
C(\bgamma)\Psi_{\bfit\gamma,E_1}\left\{
\Big(\la\!-\!E_1\!+\!\sum_{j=1}^{N-1}\!\gamma_j\Big)
\prod_{m=1}^{N-1}(\la\!-\!\gamma_m)+\sum_{j=1}^{N-1}t(\gamma_j;\bfit E)
\prod_{m\neq j} \frac{\la\!-\!\gamma_m}{\gamma_j\!-\!\gamma_m}\right\}
\ee
The expression in curly brackets is nothing but $t(\la;\bfit E)$ according
to
interpolation formula and, therefore, the function
$\Psi_{\raise-2pt\hbox{$\scriptstyle\!\!\!\bfit E $}}$ is a solution to the
spectral problem (\ref{sp3}). Note that the explicit form of
$C(\bgamma;\bfit E)$ is not required here.\square

\bigskip\noindent
{\bf Proof of Theorem \ref{th2}}. The proof is heavily based on conjecture
that all the zeros of the Hill determinant are simple
\footnote{
Up to now we are unable to prove this conjecture though it seems very
plausible.
}.
To derive the formula (\ref{gu0}) from (\ref{m1}) let us
consider the integral over $\gamma_1$ first selecting the expression
\be
\frac{c_+(\gamma_1)-\xi c_-(\gamma_1)}
{\prod\limits_{k=1}^N\sinh\frac{\textstyle\pi}{\textstyle\hbar}
\Big(\gamma_1-\delta_k\Big)}
\ee
By construction, it is an entire function but the expressions
\be
\frac{c_\pm(\gamma_1)}
{\prod\limits_{k=1}^N\sinh\frac{\textstyle\pi}{\textstyle\hbar}
\Big(\gamma_1-\delta_k\Big)}
\ee
has an infinite number of poles. Clearly, the original integral can be
written as follows while selecting the variable $\gamma_1$ and denoting
the rest of variables as $\bgamma'$:
\be\label{ff}
\int\limits_{-\infty}^\infty d\gamma_1 d\bgamma'
\frac{c_+(\gamma_1)-\xi c_-(\gamma_1)}
{\prod\limits_{k=1}^N\sinh\frac{\textstyle\pi}{\textstyle\hbar}
\Big(\gamma_1-\delta_k\Big)}
\mu^{-1}(\bgamma)\,C(\bgamma')\,\Psi_{\bgamma}=\\
=\int\limits_{{\cal C}_+}d\gamma_1d\bgamma' \,
\frac{c_+(\gamma_1)\mu^{-1}(\bgamma)}
{\prod\limits_{k=1}^N\sinh\frac{\textstyle\pi}{\textstyle\hbar}
\Big(\gamma_1-\delta_k\Big)}\,C(\bgamma')\Psi_\bgamma+\xi
\int\limits_{{\cal C}_-}d\gamma_1d\bgamma'\,
\frac{c_-(\gamma_1)\mu^{-1}(\bgamma)}
{\prod\limits_{k=1}^N\sinh\frac{\textstyle\pi}{\textstyle\hbar}
\Big(\gamma_1-\delta_k\Big)}\,C(\bgamma')\,\Psi_\bgamma
\ee
where the contour ${\cal C}_+$ encloses the poles
$\delta_k+i\hbar n_k,\;(n_k\geq 0\,,\; k=1,\ldots, N)$ while the contour
${\cal C}_-$ encloses the poles
$\delta_k+i\hbar n_k,\;(n_k< 0\,,\; k=1,\ldots, N)$.
Such deformations of contours are possible since the integrands in the r.h.s
of (\ref{ff}) vanish exponentially on large upper and
lower semi-circles respectively; this follows from the asymptotics
(\ref{as5}) and (\ref{as6}). Now the integrals can be calculated
using the residue formula. Note that $\xi c_-(\delta_k+i\hbar n_k)=
c_+(\delta_k+i\hbar n_k)$ according to (\ref{q1}). Hence,
the final expression can be written in terms of the
function $c_+$ only. The same is true for the rest of integration.\\
Let us consider the contribution coming from the
residues $\delta_1+i\hbar n_1$ while integrating over $\gamma_1$.
Using the explicit expression for the measure
\be
\mu^{-1}(\bgamma)\sim \prod_{j>k}(\gamma_j\!-\!\gamma_k)
\sinh\frac{\pi}{\hbar}(\gamma_j\!-\!\gamma_k)
\ee
one obtains the following contribution:
\be\label{d1}
\!\!\sim\!\sum_{n_1\in\z}\!c_+(\delta_1\!+\!i\hbar n_1)\!\int\!\!d\bgamma'
\prod\limits_{k=2}^{N-1}(\gamma_k\!-\!\delta_1\!-\!i\hbar n_1)
\sinh\frac{\textstyle\pi}{\textstyle\hbar}
\Big(\gamma_k-\delta_1\Big)\mu^{-1}(\bgamma')
C(\bgamma')\Psi_{\delta_1+i\hbar n_1,\gamma_2,\ldots,\gamma_{N-1}}
\ee
where we drop the common factor
\be
\left\{(N\!-\!1)!
\prod\limits_{k=2}^N\sinh\frac{\textstyle\pi}{\textstyle\hbar}
\Big(\delta_1\!-\!\delta_k\Big)\right\}^{-1}
\ee
for a moment.
Now we consider the analogous integration over variable $\gamma_2$. It is
easy
to see that there are no poles at $\gamma_2=\delta_1+i\hbar m_k$ because of
the factor $\sinh\frac{\pi}{\hbar}(\gamma_2\!-\!\delta_1)$ in
the integrand in (\ref{d1}). Quite similarly,
while calculating the residues at $\gamma_2=\delta_2+i\hbar n_2$, one
concludes that the integration over $\gamma_3$ does not include the
contribution from the points $\delta_1+i\hbar m_k$ and $\delta_2+i\hbar s_k$
and so on. Therefore, the successive evaluation of the residues at "well
arranged" points $\gamma_k=\delta_k+i\hbar n_k,\;(k=1,\ldots, N\!-\!1)$
gives the contribution
\be\label{d2}
\sim\sum_{n_1,\ldots,n_{N-1}\in\z}\prod_{k=1}^{N-1}
c_+(\delta_k\!+\!i\hbar n_k)
\Delta(\bfit\delta^{(N)}+\hbar\bfit n^{(N)})
\Psi_{\delta_1+i\hbar n_1,\ldots,\delta_{{\N}-1}+i\hbar n_{{\N}-1}}
\ee
where $\Delta(\bgamma)$ is the Vandermonde determinant and the notation
$\bfit y^{(s)}$ means the $N-1$ dimensional vector obtained from the
corresponding $N$ dimensional one by cancellation of the $s$-th component,
i.e.$\bfit y^{(s)}\equiv(y_1,\ldots,y_{s-1},y_{s+1},\ldots,y_{\N})$.
After a little algebra one finds that the common factor to (\ref{d2}) is
\be\label{d3}
(-1)^{N(N-1)/2}\left\{(N\!-\!1)!
\prod_{j>k}\sinh\frac{\pi}{\hbar}(\delta_j\!-\!\delta_k)\right\}^{-1}
\ee
By construction the integrand in (\ref{m1}) is symmetric under the
permutation of $\gamma$-variables. Therefore, one obtains the same answer
(\ref{d2}) while calculating all other possible residues from the set
$\bfit\delta^{(N)}+i\hbar\bfit n^{(N)}$. This gives the additional factor
$(N\!-\!1)!$.\\
The last step is to take into account the poles at
$\delta_{\N}+i\hbar n_{\N}$. One can easy to show that the calculation of
the corresponding residues while integrating over $\gamma_s$ gives the
additional factor $(-1)^{N-s}$ to compare with (\ref{d3}). Hence, one
obtains
the final answer (up to unessential numerical constant)
\be
\Psi_{\raise-3pt\hbox{$\scriptstyle\!\!\bfit E $}}(x_0,\bfit x)=
\frac{1}{\prod\limits_{j>k}
\sinh\frac{\pi}{\hbar}(\delta_j\!-\!\delta_k)}\times\\ \times
\sum_{s=1}^N(-1)^{N-s}\!\!\sum_{\bfit n^{(s)}\in\z^{\N-1}}
\!\!\Delta(\bfit\delta^{(s)}\!+\!i\hbar\bfit n^{(s)})\,
C_+(\bfit\delta^{(s)}\!+\!i\hbar\bfit n^{(s)})\,
\Psi_{\bfit\delta^{(s)}+i\hbar\bfit n^{(s)},E_1}(x_0,\bfit x)
\ee
where
\be
C_+(\bgamma)\equiv \prod_{k=1}^{N-1}c_+(\gamma_k)
\ee
and Theorem \ref{th2} is proved.\square

\section*{Acknowledgments}
We are deeply indebted to E. Frenkel, S. Khoroshkin, M.
Semenov-Tian-Shansky and F. Smirnov for stimulating discussions.
D.Lebedev would like to thank Mathematical Department of Berkeley University
and L.P.T.H.E., Universit\'e Pierre \'et Marie Curie for hospitality, where
the work was partially done.
The research was partly supported by grants INTAS 97-1312; RFFI
98-01-00344 (S. Kharchev); Sloan Foundation,
C.N.R.S. grant PICS No.608 and RFFI 98-01-22033 (D.Lebedev) and by grant
96-15-96455 for Support of Scientific Schools.

\end{document}